\documentclass[12pt,draftclsnofoot, onecolumn]{IEEEtran}
\usepackage{graphicx,cite,epsfig,amssymb,amsmath}
\usepackage{pifont}
\usepackage{stmaryrd}
\usepackage{bbding}
\usepackage{subfigure}
\usepackage{setspace}
\usepackage{color}
\usepackage{slashbox}
\usepackage{bm}
\usepackage{algorithm,algorithmic}
\usepackage{subeqnarray}
\allowdisplaybreaks[4]

\begin{document}
\begin{spacing}{1.65}

\title{Exploiting Inter-User Interference for Secure Massive Non-Orthogonal Multiple Access}
\author{\normalsize{Xiaoming Chen, Zhaoyang Zhang, Caijun Zhong, Derrick Wing Kwan Ng, and Rundong Jia
\thanks{Xiaoming~Chen ({\tt chen\_xiaoming@zju.edu.cn}), Zhaoyang Zhang ({\tt ning\_ming@zju.edu.cn}), Caijun Zhong ({\tt caijunzhong@zju.edu.cn}), and Rundong Jia ({\tt 3130103623@zju.edu.cn}) are with the College of Information Science and Electronic Engineering, Zhejiang University, Hangzhou, China. Derrick Wing Kwan Ng ({\tt w.k.ng@unsw.edu.au}) is with the School of Electrical Engineering and Telecommunications, the University of New South Wales, NSW, Australia.}}}\maketitle

\begin{abstract}
This paper considers the security issue of the fifth-generation (5G) wireless networks with massive connections, where multiple eavesdroppers aim to intercept the confidential messages through active eavesdropping. To realize secure massive access, non-orthogonal channel estimation (NOCE) and non-orthogonal multiple access (NOMA) techniques are combined to enhance the signal quality at legitimate users, while the inter-user interference is harnessed to deliberately confuse the eavesdroppers even without exploiting artificial noise (AN). We first analyze the secrecy performance of the considered secure massive access system and derive a closed-form expression for the ergodic secrecy rate. In particular, we reveal the impact of some key system parameters on the ergodic secrecy rate via asymptotic analysis with respect to a large number of antennas and a high transmit power at the base station (BS). Then, to fully exploit the inter-user interference for security enhancement, we propose to optimize the transmit powers in the stages of channel estimation and multiple access. Finally, extensive simulation results validate the effectiveness of the proposed secure massive access scheme.
\end{abstract}

%\begin{keywords}
%Physical layer security, massive connections, NOMA, power allocation, 5G.
%\end{keywords}

\IEEEpeerreviewmaketitle

\section{Introduction}
With the increasing development of cellular internet-of-things (IoT), the fifth-generation (5G) wireless networks are expected to support massive connections over limited radio spectrum \cite{5G1}, \cite{5G2}. Thus, it is required for 5G to adopt a novel multiple access technology to realize spectral-efficient massive access. In this context, non-orthogonal multiple access (NOMA), with a great potential of improving the spectral efficiency and admitting a large number of users simultaneously, has been widely recognized as a promising technology for 5G \cite{NOMA1,NOMA3}. On the other hand, since a massive number of heterogeneous user equipments share the same spectrum, wireless security becomes a critical issue of 5G due to the broadcast nature of wireless channels \cite{SecrecyIoT}. As such, it is desired to design a secure massive NOMA scheme.

From an information-theoretic viewpoint, in order to improve the communication secrecy performance, it is necessary to enhance the received signal quality at the legitimate users and to reduce the signal leakage to the unintended users \cite{PLS1}, \cite{PLS2}. Since the principle of NOMA is to separate the users in the power domain, power allocation is naturally applied into NOMA to enhance the wireless security. In \cite{PA1} and \cite{PA2}, power allocation schemes were designed for multiuser secure NOMA systems from the perspectives of maximizing the secrecy rate and minimizing the total power consumption, respectively. For the sake of reducing the computational complexity, a fixed-proportion power allocation scheme was proposed in \cite{NOMA2}. However, performing power allocation only has a limited capability of security enhancement, since increasing the transmit power may not only improve the signal quality at the legitimate user, but also at the eavesdropper. To effectively improve the performance of physical layer security (PHY-security), a commonly used approach is to equip with multiple antennas at the base station (BS) \cite{MIMONOMA0,MIMONOMA1,MIMONOMA10}. Then, by performing spatial beamforming at the multiple-antenna BS based on the available channel state information (CSI), it is possible to simultaneously increase the rate of the legitimate channel and decrease the rate of the eavesdropping channel, and thus improve the secrecy rate \cite{MIMONOMA2,MIMONOMA3,MIMONOMA4}. For instance, if the confidential signal is transmitted in the null space of the eavesdropping channel, the eavesdropper is unable to receive any signal. Moreover, by making use of the spatial degrees of freedom offered by the multiple antennas, artificial noise (AN) can be sent together with the desired signal to confuse the eavesdropper \cite{AN1,AN2}. Thus, in the adverse case of short-distance interception, the eavesdropper would suffer from a strong interference signal. However, in the context of massive access, there might be not enough degrees of freedom for combating the eavesdropping and transmitting the AN. To solve this problem, there are two feasible solutions. First, the required spatial degrees of freedom for the transmission of the legitimate signals can be effectively reduced via user clustering \cite{Clustering0}. For example, the users are partitioned into several clusters, and the users in a cluster share a beam \cite{Clustering1}, \cite{Clustering2}. Second, massive multiple-input multiple-output (MIMO) can be applied to the NOMA systems to significantly increase the available spatial degrees of freedom \cite{Massive1}. Especially, a BS equipped with a massive antenna array can generate high-resolution spatial beams. Thereby, the signal leakage between the users can be reduced significantly, resulting in a low inter-user interference \cite{Massive20} and a low eavesdropping channel rate \cite{Massive2}. Moreover, there are more spatial degrees of freedom to send the AN to facilitate communication secrecy provisioning \cite{Massive3}, \cite{Massive4}.

To exploit the benefits of massive MIMO for secure massive NOMA systems, the BS should obtain accurate CSI of the downlink channels \cite{CSI1}. However, it is not a trivial task for the BS to acquire accurate CSI in massive MIMO systems operated in frequency division duplex (FDD) mode or time division duplex (TDD) mode. First, in FDD mode, the amount of CSI feedback in massive MIMO systems is unbearable, since the number of required feedback bits should increase linearly proportionally to the number of BS antennas for guaranteeing a given CSI accuracy \cite{CSI2}. Thus, massive MIMO systems are usually suggested to operate in TDD mode. Specifically, by exploiting channel reciprocity, the BS obtains CSI through channel estimation based on the training sequences sent by the users over the uplink channels. However, in secure massive NOMA systems based on TDD mode, there also exists several challenging issues in CSI acquisition. On the one hand, the length of training sequence should be larger than the number of users \cite{CSI3}. In the case of massive access, the required training sequence might be longer than the duration of a channel coherence time, such that the estimated CSI has been already outdated. To effectively reduce the length of training sequence, multiple users, e.g., the users in a cluster for NOMA, should share the same training sequence. In other words, such a non-orthogonal channel estimation method reduces the length of pilot sequence at the expense of sacrificing the CSI accuracy. On the other hand, in secure communications, the eavesdroppers may actively attack the channel estimation by transmitting the same training sequence as the legitimate users \cite{CSI4}, \cite{CSI5}. As a result, the CSI accuracy is further deteriorated. With inaccurate CSI, the conventional secure communication schemes based on the AN are inapplicable, since the AN may severely interfere with the legitimate users.

In the context of massive access, especially in the presence of active attacking from the eavesdroppers, there are many challenging issues in the design of secure NOMA scheme. This is mainly because there is severe inter-user interference during the stages of CSI acquisition and multiple access. The inter-user interference decreases the CSI accuracy and the quality of the legitimate signal, resulting in a high probability of interception. However, the inter-user interference also has a great impact on the quality of the eavesdropping signal. Specifically, the inter-user interference can be harnessed and served as AN signal to confuse the eavesdroppers. Inspired by that, this paper proposes to exploit the inter-user interference to enhance the secrecy performance of secure massive NOMA systems. The contributions of this paper are three-fold:

\begin{enumerate}

\item We design a secure non-orthogonal transmission framework for 5G wireless networks with massive connections, where the originally harmful inter-user interference is exploited to enhance the communication secrecy performance.

\item We analyze the performance of the proposed secure massive NOMA system, and derive a closed-form expression for the ergodic secrecy rate. Moreover, we reveal the relationship between ergodic secrecy rate and various important system parameters via asymptotic analysis.

\item We propose two optimization schemes for the design of channel estimation and multiple access of secure massive NOMA systems, so as to fully exploit the benefits of inter-user interference for security enhancement.

\end{enumerate}

The rest of this paper is outlined as follows. Section II designs a non-orthogonal transmission framework for secure massive NOMA systems. Section III focuses on the analysis of the secrecy performance of the secure massive NOMA system, and reveals the relationship between ergodic secrecy rate and system parameters. Section IV proposes two power allocation schemes at the user and the BS to fully exploit the inter-user interference. Section V presents extensive simulation results to validate the effectiveness of the proposed schemes, finally Section VI concludes the whole paper.

\emph{Notations}: We use bold upper (lower) letters to denote matrices (column vectors), $(\cdot)^H$ to denote conjugate transpose, $\mathrm{E}[\cdot]$ to denote expectation, $\mathrm{var}(\cdot)$ to denote the variance, $\|\cdot\|$ to denote the $L_2$-norm of a vector, $\otimes$ to denote the Kronecker product, $\mathrm{vec}(\cdot)$ to denote the vectorization of a matrix, $|\cdot|$ to denote the absolute value, $\chi^2(N)$ to denote the $\chi^2$ distribution with $N$ degrees of freedom, and $\mathcal{CN}(x,y)$ to denote the complex Gaussian distribution with mean $x$ and variance $y$.

\section{System Model}
\begin{figure}[h] \centering
\includegraphics [width=0.65\textwidth] {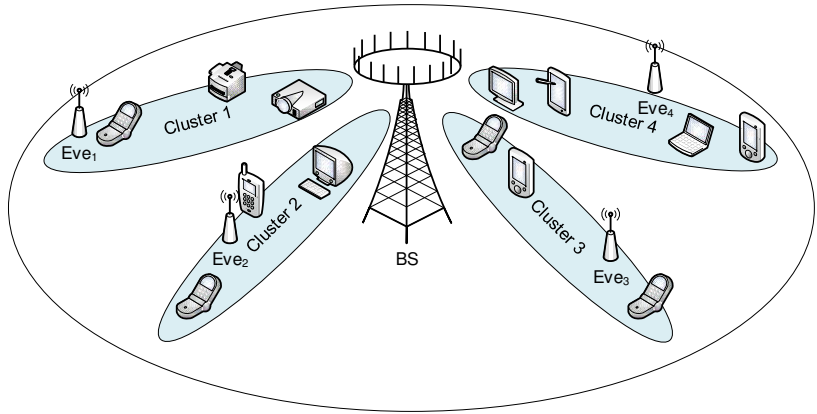}
\caption {A secure massive NOMA system with four clusters.}
\label{Fig1}
\end{figure}

Consider a single-cell multiuser downlink communication
system\footnote{The proposed schemes can be extended to the case of
multiple-cell directly. We will discuss the multi-cell case in
Section III.}, where a base station (BS) equipped with $N_t$
antennas broadcasts messages to $K$ single-antenna user equipments
(UEs), cf. Fig. \ref{Fig1}. Note that both the number of BS antennas
$N_t$ and the number of UEs $K$ might be very large in future 5G
wireless networks. In other words, we consider a massive access
cellular system by making use of the massive multiple-input
multiple-output (MIMO) technique. To reduce the computational
complexity, the $K$ UEs are grouped into $M$ clusters. In
particular, the UEs in the same direction but with distinct
propagation distances based on their spatial directions are
partitioned into a cluster for exploiting the spatial degrees of
freedom offered by the multiple-antenna BS\footnote{The spatial
direction of users can be found via various methods/technologies
such as GPS or user location tracking algorithms.}. Without loss of
generality, it is assumed that there are $N_m$ UEs in the $m$th
cluster. In each cluster, we assume that there is an eavesdropper
(Eve) intercepting the messages sent to the UEs in that cluster. In
particular, the Eve is or pretends to be a UE, thus we assume that
it is also equipped with a single antenna. For ease of notation, we
use UE$_{m,n}$ and Eve$_m$ to denote the $n$th UE and the associated
eavesdropper in the $m$th cluster, respectively. To guarantee secure
communication, PHY-security technique is applied to the massive
access system. To fully exploiting the benefits of multiple-antenna
BS for enhancing the secrecy performance, the BS should have partial
CSI of the downlink channels. Thus, we first design a CSI
acquisition method which is suitable for massive access.

\subsection{Non-Orthogonal Channel Estimation}
Similar to most related works about massive MIMO systems, the considered system operates in TDD mode for ease of CSI acquisition. To be specific, the UEs first send training sequences over the uplink channels, and then the BS obtains the CSI of the uplink channels through channel estimation. Due to the property of channel reciprocity in TDD mode, the CSI of the uplink channels can be used as the CSI of the corresponding downlink channels. For the traditional CSI acquisition methods, in order to guarantee accurate CSI, the training sequences are designed with pairwise orthogonality. However, in the context of massive access, the length of the training sequence might be longer than the channel coherence time for guaranteeing pairwise orthogonality, resulting in outdated CSI. On the other hand, for eavesdropping more information, the Eves may interfere with the channel estimation by transmitting the same training sequence as the legitimate users. To simultaneously handle the two challenging issues, we propose a non-orthogonal channel estimation method for secure massive access in the following.

At the beginning of each time slot, the UEs in the $m$th cluster simultaneously send a common training sequence $\mathbf{\Phi}_m^H$ with a length of $\tau$ symbols over the uplink channels, while the Eve$_m$ actively attacks the channel estimation through transmitting the same training sequence $\mathbf{\Phi}_m^H$ as the UEs. We note that although the training sequences in a cluster are the same, the training sequences across the clusters are pairwise orthogonal, namely $\mathbf{\Phi}_i^H\mathbf{\Phi}_j=0, \forall i\neq j,$ and $\mathbf{\Phi}_i^H\mathbf{\Phi}_i=1$. Hence, the length of the training sequences is only required to be longer than the number of clusters $M$, which usually can be realized in practical systems. Based on the non-orthogonal training sequences, the received training signal at the BS can be expressed as
\begin{equation}
\mathbf{Y}={\underbrace{\sum\limits_{j=1}^{M}\sum\limits_{i=1}^{N_j}\sqrt{\alpha_{j,i}Q_{j,i}\tau}\mathbf{h}_{j,i}\mathbf{\Phi}_j^H}_{\textrm{Legitimate training signal}}} +{\underbrace{\sum\limits_{j=1}^{M}\sqrt{\beta_{j}U_{j}\tau}\mathbf{g}_{j}\mathbf{\Phi}_j^H}_{\textrm{Attacking signal}}}+{\underbrace{\mathbf{N}}_{\textrm{AWGN}}},\label{eqn1}
\end{equation}
where $\alpha_{j,i}$ and $\mathbf{h}_{j,i}$ denote the path loss and the $N_t$-dimensional small-scale fading vector of the channel from the UE$_{j,i}$ to the BS. Variables $\beta_j$ and $\mathbf{g}_{j}$ are the path loss and the $N_t$-dimensional small-scale fading vector from the Eve$_j$ to the BS, respectively. It is assumed that $\alpha_{j,i}$ and $\beta_j$ remain constant during a relatively long time, while $\mathbf{h}_{j,i}$ and $\mathbf{g}_{j}$ independently fade over time slots according to the distribution $\mathcal{CN}(0,\mathbf{I}_{N_t})$. Moreover, $Q_{j,i}$ and $U_{j}$ represent the transmit power of the UE$_{j,i}$ and the Eve$_j$, and $\mathbf{N}$ is an $N_t\times\tau$ additive white Gaussian noise (AWGN) matrix with zero mean and unit variance elements. Since the UEs in a cluster share the same pilot sequence, we only need to estimate the effective CSI $\mathbf{h}_j$ for the $j$th cluster, where $j=1,\cdots,M$. Let us focus on the estimation of $\mathbf{h}_m$ as an example. First, right-multiplying $\mathbf{Y}$ by $\mathbf{\Phi}_m$, we have
\begin{eqnarray}
\mathbf{Y}\mathbf{\Phi}_m&=&\sum\limits_{i=1}^{N_m}\sqrt{\alpha_{m,i}Q_{m,i}\tau}\textbf{h}_{m,i}+\sqrt{\beta_{m}U_{m}\tau}\mathbf{g}_{m}+\mathbf{N}\mathbf{\Phi}_m\nonumber\\
&=&\sqrt{\sum\limits_{i=0}^{N_m}\alpha_{m,i}Q_{m,i}\tau}\mathbf{h}_m+\mathbf{N}\mathbf{\Phi}_m,\label{eqn2}
\end{eqnarray}
where $\mathbf{h}_m=\frac{\sum\limits_{n=0}^{N_m}\sqrt{\alpha_{m,n}Q_{m,n}\tau}\mathbf{h}_{m,n}}{\sqrt{\sum\limits_{i=0}^{N_m}\alpha_{m,i}Q_{m,i}\tau}}$ is the effective CSI for the $m$th cluster in presence of the active attacking signal from the Eve$_m$. Note that we let $\beta_{m}=\alpha_{m,0}$, $U_m=Q_{m,0}$, and $\mathbf{g}_m=\mathbf{h}_{m,0}$ for clarity of notation. Then, by using minimum mean square error (MMSE) channel estimation, the estimated CSI $\hat{\mathbf{h}}_{m}$ at the BS is given by
\begin{eqnarray}
\hat{\mathbf{h}}_{m}=\frac{\sqrt{\sum\limits_{n=0}^{N_m}\alpha_{m,n}Q_{m,n}\tau}}{1+\sum\limits_{i=0}^{N_m}\alpha_{m,i}Q_{m,i}\tau}\hat{\mathbf{y}}_{m},\label{eqn3}
\end{eqnarray}
where
\begin{equation}
\hat{\mathbf{y}}_{m}=\sum\limits_{i=0}^{N_m}\sqrt{\alpha_{m,i}Q_{m,i}\tau}\textbf{h}_{m,i}+(\mathbf{\Phi}_m\otimes\mathbf{I}_{N_t})\mathbf{n},\label{eqn4}
\end{equation}
and $\mathbf{n}=\textrm{vec}(\mathbf{N})$. Combining (\ref{eqn3}) and (\ref{eqn4}), for the UE$_{m,n}$, the relation between the actual CSI $\mathbf{h}_{m,n}$ and the estimated CSI $\hat{\mathbf{h}}_{m}$ can be expressed as
\begin{equation}
\mathbf{h}_{m,n}=\sqrt{\rho_{m,n}}\hat{\mathbf{h}}_{m}+\sqrt{1-\rho_{m,n}}\mathbf{e}_{m,n},\label{eqn5}
\end{equation}
where $\mathbf{e}_{m,n}$ is the channel estimation error vector with
independently and identical distribution (i.i.d.) zero mean and unit
variance complex Gaussian distributed entries, and is independent of
$\hat{\mathbf{h}}_{m}$. The variable
$\rho_{m,n}=\frac{\alpha_{m,n}Q_{m,n}\tau}{1+\sum\limits_{i=0}^{N_m}\alpha_{m,i}Q_{m,i}\tau}$
($0\leq\rho_{m,n}\leq1, \forall m,n$) is the correlation coefficient
between $\mathbf{h}_{m,n}$ and $\hat{\mathbf{h}}_{m}$. Specifically,
a large $\rho_{m,n}$ means a high CSI accuracy, which is affected by
the inter-user interference. Note that the effective CSI
$\hat{\mathbf{h}}_{m}$ also contains partial information of the
eavesdropper channel $\mathbf{g}_m$, namely $\mathbf{h}_{m,0}$. If
the Eves are passive, we have $\rho_{m,0}=0$. In other words, the
case of passive Eves is a special one of the considered model, and
thus the proposed schemes in the following is also applicable in the
case of passive Eves.

\emph{Remark 1}: Non-orthogonal channel estimation significantly shortens the length of training sequences which facilitates CSI acquisition in the scenario of massive access. Meanwhile, the proposed non-orthogonal channel estimation leads to the decrease of CSI accuracy, which is similar to pilot contamination in multi-cell massive MIMO systems. However, the impact of low accurate CSI caused by non-orthogonal channel estimation can be alleviated via successive interference cancellation (SIC) at the UEs, but the interference due to pilot contamination is difficult to be mitigated. In addition, non-orthogonal channel estimation decreases the number of required training sequences in each cell, and thus it is possible to solve the problem of pilot contamination due to pilot shortage. More importantly, it is able to fight against the active attacking from the Eves by making use of inter-user interference during non-orthogonal channel estimation. Especially, one can coordinate the transmit power $Q_{m,n}$ to further alleviate the negative impact due to active attacking. Thereby, non-orthogonal channel estimation is more appealing for secure massive NOMA systems.

\subsection{Non-Orthogonal Multiple Access}
In order to support massive access over limited radio spectrum, the considered system also adopts a non-orthogonal scheme based on user clustering during the stage of signal transmission. Specifically, the BS first performs superposition coding to the signals of the $m$th cluster as below:
\begin{equation}
x_m=\sum\limits_{n=1}^{N_m}\sqrt{P_{m,n}}s_{m,n},\label{eqn6}
\end{equation}
where $s_{m,n}$ and $P_{m,n}$ are the Gaussian distributed signal with a unit norm and the corresponding transmit power for the UE$_{m,n}$, respectively. Then, the BS combines all clusters' signals as follows:
\begin{equation}
\mathbf{x}=\sum\limits_{m=1}^{M}\mathbf{w}_mx_m,\label{eqn7}
\end{equation}
where $\mathbf{w}_m$ is an $N_t$-dimensional transmit beam designed for the $m$th cluster based on the obtained CSI through non-orthogonal channel estimation. In this paper, we adopt a low-complexity maximum ratio transmission (MRT) beamforming scheme\footnote{MRT can achieve the asymptotically optimal performance as the number of BS antennas $N_t$ approaches infinity \cite{Massive20}.}. For such a beamforming scheme, $\mathbf{w}_m$ can be expressed as
\begin{eqnarray}
\mathbf{w}_m&=&\frac{\hat{\mathbf{h}}_{m}}{\|\hat{\mathbf{h}}_{m}\|}.\label{eqn8}
\end{eqnarray}
Finally, the BS broadcasts the signal $\mathbf{x}$ over the downlink channels. Therefore, the received signal at the UE$_{m,n}$ is given by
\begin{eqnarray}
y_{m,n}^o\!\!\!&=&\!\!\!\alpha_{m,n}^{1/2}\textbf{h}_{m,n}^H\textbf{x}+n_{m,n}\nonumber\\
&=&\!\!\!{\underbrace{\alpha_{m,n}^{1/2}\textbf{h}_{m,n}^H\textbf{w}_{m}\sqrt{P_{m,n}}s_{m,n}}_{\textrm{Desired signal}}}+{\underbrace{\sum\limits_{i=1,i\neq n}^{N_m}\alpha_{m,n}^{1/2}\textbf{h}_{m,n}^H\textbf{w}_{m}\sqrt{P_{m,i}}s_{m,i}}_{\textrm{Intra-cluster interference}}}\nonumber\\
&&\!\!\!+{\underbrace{\sum\limits_{j=1,j\neq
m}^{M}\sum\limits_{i=1}^{N_j}\alpha_{m,n}^{1/2}\textbf{h}_{m,n}^H\textbf{w}_{j}\sqrt{P_{j,i}}s_{j,i}}_{\textrm{Inter-cluster
interference}}}+{\underbrace{n_{m,n}^{o}}_{\mathrm{AWGN}}},\label{eqn9}
\end{eqnarray}
where $n_{m,n}^{o}$ is the AWGN with zero mean and unit variance. Note that the
second term at the right hand of (\ref{eqn9}) is the intra-cluster
interference, which can be partially cancelled by using SIC techniques at the UEs for enhancing
the legitimate signal quality. Without loss of generality, it is
assumed that the effective channel gains related to the UEs in the
$m$th cluster have a descending order, namely
\begin{equation}
|\alpha_{m,1}^{1/2}\mathbf{h}_{m,1}^H\mathbf{w}_m|^2\geq|\alpha_{m,2}^{1/2}\mathbf{h}_{m,2}^H\mathbf{w}_m|^2\geq\cdots\geq|\alpha_{m,N_m}^{1/2}\mathbf{h}_{m,N_m}^H\mathbf{w}_m|^2.\label{eqn10}
\end{equation}
Each UE perfectly knows its effective channel gain through channel
estimation, which is conveyed to the BS over the uplink links. Then,
the BS obtains the order of effective channel gains in each cluster
and informs the UEs. As such, the UE$_{m,n}$ first decodes the
signals of the UE$_{m,i}$, $i\in[n+1,N_m]$, in the reverse order,
then subtracts the intra-interference from these UEs in its received
signal $y_{m,n}^o$, and finally demodulates its desired signal
$s_{m,n}$. In other words, the residual intra-cluster interference
is only from the UEs with stronger channel gains after
SIC\footnote{In Practice, due to channel estimation error, hardware
limitation of the UEs, the low signal quality, and other factors,
the decoding error of the weak interfering signal may occur. As a
result, there exists residual interference from the weak UEs after
SIC, namely imperfect SIC \cite{ImperfectSIC1}. We will discuss the
impact of imperfect SIC in Section III.}. Thus, for the UE$_{m,n}$,
the intra-cluster interference is reduced to
$\sum\limits_{i=1}^{n-1}\alpha_{m,n}^{1/2}\textbf{h}_{m,n}^H\textbf{w}_{m}\sqrt{P_{m,i}}s_{m,i}$,
and its received signal-to-interference-plus-noise ratio (SINR) can
be computed as
\begin{eqnarray}
\gamma_{m,n}^o=\frac{|\mathbf{h}_{m,n}^H\mathbf{w}_m|^2\alpha_{m,n}P_{m,n}}{|\mathbf{h}_{m,n}^H\mathbf{w}_m|^2\alpha_{m,n}\sum\limits_{i=1}^{n-1}P_{m,i}+\sum\limits_{j=1,j\neq m}^{M}\sum\limits_{i=1}^{N_j}|\mathbf{h}_{m,n}^H\mathbf{w}_j|^2\alpha_{m,n}P_{j,i}+1}.\label{eqn11}
\end{eqnarray}
Meanwhile, the Eves also receive the signal from the BS. The eavesdropping signal related to $s_{m,n}$ and the corresponding SINR at the Eve$_m$ can be expressed as
\begin{eqnarray}
y_{m,n}^e\!\!\!&=&\!\!\!{\underbrace{\beta_{m}^{1/2}\textbf{g}_{m}^H\textbf{w}_{m}\sqrt{P_{m,n}}s_{m,n}}_{\textrm{Desired signal}}}+{\underbrace{\sum\limits_{i=1,i\neq n}^{N_m}\beta_{m}^{1/2}\textbf{g}_{m}^H\textbf{w}_{m}\sqrt{P_{m,i}}s_{m,i}}_{\textrm{Intra-cluster interference}}}\nonumber\\
&&\!\!\!+{\underbrace{\sum\limits_{j=1,j\neq
m}^{M}\sum\limits_{i=1}^{N_j}\beta_{m}^{1/2}\textbf{g}_{m}^H\textbf{w}_{j}\sqrt{P_{j,i}}s_{j,i}}_{\textrm{Inter-cluster
interference}}}+{\underbrace{n_{m}^{e}}_{\mathrm{AWGN}}},\label{eqn12}
\end{eqnarray}
and
\begin{eqnarray}
\gamma_{m,n}^e=\frac{|\mathbf{g}_{m}^H\mathbf{w}_m|^2\beta_{m}P_{m,n}}{|\mathbf{g}_{m}^H\mathbf{w}_m|^2\beta_{m}\sum\limits_{i=1,i\neq n}^{N_m}P_{m,i}+\sum\limits_{j=1,j\neq m}^{M}\sum\limits_{i=1}^{N_j}|\mathbf{g}_{m}^H\mathbf{w}_j|^2\beta_{m}P_{j,i}+1},\label{eqn13}
\end{eqnarray}
where $n_{m}^{e}$ is the AWGN with zero mean and unit variance at
the Eve$_m$. Note that the Eves do not perform SIC within a cluster,
since it is difficult for the Eves to know the ordering of the
effective channel gains, namely the reverse ordering of SIC. If the
Eves Perform SIC with an arbitrary ordering, the decoding error
might be inevitable. Moreover, through exploiting the inter-user
interference, it is possible to make the eavesdropping channel rate
less than a given threshold, which is not larger than the transmit
rate of the confidential signals. In this case, if the Eves try to
perform SIC, the decoding error will also occur since the
eavesdropping channel rate is less than the transmit rate. In fact,
if the Eves are capable of decoding the weak signals for SIC, secure
communication can not be realized for a ceratin UEs at least.

\emph{Remark 2}: Due to non-orthogonal multiple access, inter-user interference exists inevitably, resulting in a potential performance degradation. However, the originally harmful inter-user interference can be used to confuse the Eves, and thus enhances the communication security. In this paper, we aim to exploit the inter-user interference for balancing the interference to the UEs and to the Eves, so as to improve the secrecy performance of secure massive NOMA systems.

\section{Secrecy Performance Analysis of Massive NOMA}
In this section, we study the ergodic secrecy rate of secure massive NOMA communications and reveal the impacts of inter-user interference caused by non-orthogonal channel estimation and non-orthogonal multiple access. The results of this section shed lights on the optimization of system parameters for enhancing the secrecy performance.

\subsection{Ergodic Secrecy Rate}
In general, the ergodic secrecy rate of the UE$_{m,n}$ can be computed as
\begin{equation}
R_{m,n}^{sec}=\mathrm{E}[(R_{m,n}^{o}-R_{m,n}^{e})^{+}],\label{eqn14}
\end{equation}
where $R_{m,n}^{o}$ and $R_{m,n}^{e}$ are the rates of the legitimate channel and the eavesdropping channel related to the UE$_{m,n}$, respectively. In the case that the BS is equipped with a massive number of antennas, the ergodic secrecy rate is lower bounded by \cite{Massive3}
\begin{equation}
\bar{R}_{m,n}^{sec}=(\bar{R}_{m,n}^{o}-\bar{R}_{m,n}^{e})^{+},\label{eqn15}
\end{equation}
where $\bar{R}_{m,n}^{o}=\mathrm{E}[R_{m,n}^{o}]$ and $\bar{R}_{m,n}^{e}=\mathrm{E}[R_{m,n}^{e}]$ are the ergodic rates of the legitimate channel and the eavesdropping channel, respectively. In other words, one can obtain a lower bound on the ergodic secrecy rate by computing the achievable rate of the legitimate channel and the eavesdropping channel. In what follows, we analyze the two ergodic rates in detail, respectively. Firstly, according to the definition of the received SINR of the UE$_{m,n}$ in (\ref{eqn11}), the ergodic rate of the legitimate channel can be expressed as
\begin{eqnarray}
\bar{R}_{m,n}^{o}=\mathrm{E}[\log_2(1+\gamma_{m,n}^{o})].\label{eqn16}
\end{eqnarray}
However, it is a nontrivial task to compute the expectation with respect to a complicated random variable $\gamma_{m,n}^{o}$. To solve this challenge, we resort to an approximation of the ergodic rate of the legitimate channel. To be more specific, we rewrite the after-SIC signal at the UE$_{m,n}$ as
\begin{equation}
y_{m,n}^o=I_{m,n,0}+n_{m,n}^{'},\label{eqn17}
\end{equation}
where $I_{m,n,0}=\mathrm{E}[\alpha_{m,n}^{1/2}\textbf{h}_{m,n}^H\textbf{w}_{m}\sqrt{P_{m,n}}s_{m,n}]$ is the mean of the desired signal, and $n_{m,n}^{'}$ is the effective noise, which is given by
\begin{equation}
n_{m,n}^{'}=\sum\limits_{t=1}^{3}I_{m,n,t}+n_{m,n}^o,\label{eqn18}
\end{equation}
where $I_{m,n,1}=\alpha_{m,n}^{1/2}\textbf{h}_{m,n}^H\textbf{w}_{m}\sqrt{P_{m,n}}s_{m,n}-\mathrm{E}[\alpha_{m,n}^{1/2}\textbf{h}_{m,n}^H\textbf{w}_{m}\sqrt{P_{m,n}}s_{m,n}]$ is the signal leakage, $I_{m,n,2}=\sum\limits_{i=1}^{n-1}\alpha_{m,n}^{1/2}\textbf{h}_{m,n}^H\textbf{w}_{m}\sqrt{P_{m,i}}s_{m,i}$ is the residual intra-cluster interference, and $I_{m,n,3}=\sum\limits_{j=1,j\neq
m}^{M}$ $\sum\limits_{i=1}^{N_j}\alpha_{m,n}^{1/2}\textbf{h}_{m,n}^H\textbf{w}_{j}\sqrt{P_{j,i}}s_{j,i}$ is the inter-cluster interference, respectively. If $n_{m,n}^{'}$ is the AWGN independent of $I_{m,n,0}$ and the number of BS antennas is large, we can obtain an approximated ergodic rate as \cite{Lowerbound}
\begin{eqnarray}
\bar{R}_{m,n}^{o}\approx\log_2(1+\tilde{\gamma}_{m,n}),\label{eqn19}
\end{eqnarray}
where $\tilde{\gamma}_{m,n}=\frac{\phi_{m,n,0}}{\sum\limits_{t=1}^{3}\phi_{m,n,t}+1}$ is the effective SINR with
\begin{equation}
\phi_{m,n,0}=|I_{m,n,0}|^2,\label{eqn20}
\end{equation}
and
\begin{equation}
\phi_{m,n,t}=\mathrm{E}[|I_{m,n,t}|^2],t=1,\dots,3,\label{eqn21}
\end{equation}
being the powers of the desired signal, the signal leakage, the intra-cluster interference, and the inter-cluster interference, respectively. Thus, the derivation of the ergodic rate of the legitimate channel is equivalent to the computation of expectations in (\ref{eqn20}) and (\ref{eqn21}). In what follows, we compute these expectations in detail, respectively.

We first consider the term $I_{m,n,0}$ for the computation of $\phi_{m,n,0}$. Substituting $\mathbf{w}_m=\frac{\hat{\mathbf{h}}_{m}}{\|\hat{\mathbf{h}}_{m}\|}$ and $\mathbf{h}_{m,n}=\sqrt{\rho_{m,n}}\hat{\mathbf{h}}_{m}+\sqrt{1-\rho_{m,n}}\mathbf{e}_{m,n}$ into $I_{m,n,0}$, we have
\begin{subequations}
\begin{align}
I_{m,n,0}&=\sqrt{\alpha_{m,n}P_{m,n}}\mathrm{E}[\sqrt{\rho_{m,n}}\hat{\mathbf{h}}_{m}^H\mathbf{w}_m+\sqrt{1-\rho_{m,n}}\mathbf{e}_{m,n}^H\mathbf{w}_m]\\
&=\sqrt{\alpha_{m,n}P_{m,n}\rho_{m,n}}\mathrm{E}[\hat{\mathbf{h}}_{m}^H\mathbf{w}_m]\label{eqn22}\\
&=\sqrt{\alpha_{m,n}P_{m,n}\rho_{m,n}}\mathrm{E}[\|\hat{\mathbf{h}}_{m}\|]\\
&=\sqrt{\alpha_{m,n}P_{m,n}\rho_{m,n}}\frac{\Gamma(N_t+\frac{1}{2})}{\Gamma(N_t)},\label{eqn23}
\end{align}
\end{subequations}
where $\Gamma(\cdot)$ denotes the Gamma function. Eq. (\ref{eqn22}) follows the fact of $\mathrm{E}[\mathbf{e}_{m,n}^H\mathbf{w}_m]=\mathrm{E}[\mathbf{e}_{m,n}^H]\mathrm{E}[\mathbf{w}_m]$ $=0$, and Eq. (\ref{eqn23}) holds true since $\|\hat{\mathbf{h}}_{m}\|$ has a scaled $\chi^2(2N_t)$ distribution by a factor $\frac{1}{2}$. Equivalently, we have $\mathrm{E}[\|\mathbf{h}_{m}\|]=\frac{\Gamma(N_t+\frac{1}{2})}{\Gamma(N_t)}$. According to (\ref{eqn23}), it is easy to obtain the power of the desired signal as
\begin{equation}
\phi_{m,n,0}=\alpha_{m,n}P_{m,n}\rho_{m,n}\frac{\Gamma^2(N_t+\frac{1}{2})}{\Gamma^2(N_t)}.\label{eqn24}
\end{equation}
Furthermore, if the number of BS antennas $N_t$ is large, the power of the desired signal can be approximated as
\begin{equation}
\phi_{m,n,0}\approx\alpha_{m,n}P_{m,n}\rho_{m,n}N_t,\label{eqn25}
\end{equation}
where Eq. (\ref{eqn25}) follows the fact of $\frac{\Gamma^2(N_t+\frac{1}{2})}{\Gamma^2(N_t)}\rightarrow N_t$ when $N_t$ is large. It is seen that the CSI accuracy $\rho_{m,n}$ has a direct impact on the power of the desired signal.

Then, we analyze the power of the signal leakage, which can be computed as
\begin{subequations}
\begin{align}
\phi_{m,n,1}&=\mathrm{E}\left[|I_{m,n,1}|^2\right]\\
&=\mathrm{var}\left[\alpha_{m,n}^{1/2}\textbf{h}_{m,n}^H\textbf{w}_{m}\sqrt{P_{m,n}}s_{m,n}\right]\\
&=\mathrm{E}\left[\left(\alpha_{m,n}^{1/2}\textbf{h}_{m,n}^H\textbf{w}_{m}\sqrt{P_{m,n}}\right)^2\right]-\mathrm{E}^2\left[\alpha_{m,n}^{1/2}\textbf{h}_{m,n}^H\textbf{w}_{m}\sqrt{P_{m,n}}\right]\\
&=\alpha_{m,n}P_{m,n}\left(\mathrm{E}\left[|\textbf{h}_{m,n}^H\textbf{w}_{m}|^2\right]-\mathrm{E}^2\left[\textbf{h}_{m,n}^H\textbf{w}_{m}\right]\right)\\
&=\alpha_{m,n}P_{m,n}\left(\mathrm{E}\left[|\sqrt{\rho_{m,n}}\hat{\mathbf{h}}_{m}^H\textbf{w}_{m}|^2\right]+\mathrm{E}\left[|\sqrt{1-\rho_{m,n}}\mathbf{e}_{m,n}^H\textbf{w}_{m}|^2\right]-\mathrm{E}^2\left[\textbf{h}_{m,n}^H\textbf{w}_{m}\right]\right)\label{eqn26}\\
&=\alpha_{m,n}P_{m,n}\left(\rho_{m,n}N_t+1-\rho_{m,n}-\rho_{m,n}\frac{\Gamma^2(N_t+\frac{1}{2})}{\Gamma^2(N_t)}\right),\label{eqn27}
\end{align}
\end{subequations}
where Eq. (\ref{eqn26}) holds true since $\hat{\mathbf{h}}_{m}$ and $\mathbf{e}_{m,n}$ are independent of each other and Eq. (\ref{eqn27}) follows the facts of $\mathrm{E}\left[\|\hat{\mathbf{h}}_{m}\|^2\right]=N_t$ and $\mathrm{E}\left[|\mathbf{e}_{m,n}^H\textbf{w}_{m}|^2\right]=1$. Also, if the number of BS antennas is sufficiently large, the term $1-\rho_{m,n}$ in (\ref{eqn27}) is negligible, such that the power of the signal leakage can be approximated as
\begin{equation}
\phi_{m,n,1}\approx0.\label{eqn28}
\end{equation}
Intuitively, the signal leakage is caused by channel fading. However, the downlink channels becomes deterministic due to channel hardening as $N_t$ tends to infinity, and thus the power of the signal leakage asymptotically approaches zero.

Similarly, it is easy to derive the powers of the residual intra-cluster interference and the inter-cluster interference as
\begin{subequations}
\begin{align}
\phi_{m,n,2}&=\mathrm{E}\left[|I_{m,n,2}|^2\right]\\
&=\alpha_{m,n}\mathrm{E}\left[|\textbf{h}_{m,n}^H\textbf{w}_{m}|^2\right]\sum\limits_{i=1}^{n-1}P_{m,i}\label{eqn29}\\
&=\alpha_{m,n}\left(\mathrm{E}\left[|\sqrt{\rho_{m,n}}\hat{\mathbf{h}}_{m}^H\textbf{w}_{m}|^2\right]+\mathrm{E}\left[|\sqrt{1-\rho_{m,n}}\mathbf{e}_{m,n}^H\textbf{w}_{m}|^2\right]\right)\sum\limits_{i=1}^{n-1}P_{m,i}\label{eqn30}\\
&=\alpha_{m,n}\left(\rho_{m,n}N_t+1-\rho_{m,n}\right)\sum\limits_{i=1}^{n-1}P_{m,i}\label{eqn31}\\
&\approx\alpha_{m,n}\rho_{m,n}N_t\sum\limits_{i=1}^{n-1}P_{m,i},\label{eqn32}
\end{align}
\end{subequations}
and
\begin{subequations}
\begin{align}
\phi_{m,n,3}&=\mathrm{E}\left[|I_{m,n,3}|^2\right]\\
&=\sum\limits_{j=1,j\neq
m}^{M}\sum\limits_{i=1}^{N_j}\alpha_{m,n}P_{j,i}\mathrm{E}\left[|\textbf{h}_{m,n}^H\textbf{w}_{j}|^2\right]\label{eqn33}\\
&=\alpha_{m,n}\sum\limits_{j=1,j\neq
m}^{M}\sum\limits_{i=1}^{N_j}P_{j,i},\label{eqn34}
\end{align}
\end{subequations}
respectively, where Eq. (\ref{eqn29}) and (\ref{eqn33}) hold true since the signals $s_{j,i}$ are independent of each other, Eq. (\ref{eqn30}) follows the fact that $\hat{\mathbf{h}}_{m}$ and $\mathbf{e}_{m,n}$ are independent of each other, Eq. (\ref{eqn32}) neglects the term $1-\rho_{m,n}$, and Eq. (\ref{eqn34}) is obtained because $|\textbf{h}_{m,n}^H\textbf{w}_{j}|^2$ has a scaled $\chi^2(2)$ distribution by a factor $\frac{1}{2}$. Substituting (\ref{eqn25}), (\ref{eqn27}), (\ref{eqn32}), and (\ref{eqn34}) into (\ref{eqn19}), the ergodic rate of the legitimate channel in secure massive NOMA systems can be expressed as
\begin{eqnarray}
\bar{R}_{m,n}^{o}\approx\log_2\left(1+\frac{\alpha_{m,n}P_{m,n}\rho_{m,n}N_t}{\alpha_{m,n}\rho_{m,n}N_t\sum\limits_{i=1}^{n-1}P_{m,i}+\alpha_{m,n}\sum\limits_{j=1,j\neq
m}^{M}\sum\limits_{i=1}^{N_j}P_{j,i}+1}\right).\label{eqn35}
\end{eqnarray}

Now, we analyze the ergodic rate of the eavesdropping channel, $\bar{R}_{m,n}^{e}$. It is noticed that the Eve can be regarded as a special UE without performing SIC. Thus, similar to the derivation of $\bar{R}_{m,n}^{o}$, $\bar{R}_{m,n}^{e}$ can be expressed as
\begin{eqnarray}
\bar{R}_{m,n}^{e}&=&\mathrm{E}[1+\gamma_{m,n}^e]\nonumber\\
&\approx&\log_2\left(1+\frac{\beta_{m}P_{m,n}\rho_{m,0}N_t}{\beta_{m}\rho_{m,0}N_t\sum\limits_{i=1,i\neq n}^{N_m}P_{m,i}+\beta_{m}\sum\limits_{j=1,j\neq
m}^{M}\sum\limits_{i=1}^{N_j}P_{j,i}+1}\right).\label{eqn36}
\end{eqnarray}
Note that if the Eves do not actively attack the channel estimation
by transmitting the same training sequence, they cannot receive any
of the desired signal due to the fact $\rho_{m,0}=0$. Moreover, if
the Eves only send the interference signal independent of the
training sequence, they are also unable to intercept the message.
This is because the massive antenna array can generate
high-resolution spatial beams to avoid the signal leakage to the
unintended users. However, massive MIMO systems operating in TDD
mode is preferable, therefore it is vulnerable to active
eavesdropping during channel estimation.

Substituting (\ref{eqn35}) and (\ref{eqn36}) into (\ref{eqn15}), we
can obtain a closed-form expression for a lower bound on the ergodic
secrecy rate $\bar{R}_{m,n}^{sec}$. It is found that
$\bar{R}_{m,n}^{sec}$ is a function of the CSI accuracy $\rho_{m,n}$
and the transmit power $P_{m,n}$. The derived lower bound can be
regarded as the case of channel hardening. Thus, when the number of
BS antennas is sufficiently large, the lower bound is tight. In
fact, as seen in Fig. \ref{Fig2}, even with a finite number of BS
antennas, e.g., $N_t=64$, the lower bound is accurate. Notice that
$\rho_{m,n}$ is determined by non-orthogonal channel estimation, and
$P_{m,n}$ is an important parameter for controlling the interference
caused by non-orthogonal multiple access. As such, it is possible to
improve the secrecy performance by optimizing these system
parameters, which is equivalent to coordinating the inter-user
interference to enhance the communication secrecy. Moreover, it is
seen that the number of BS antennas $N_t$ also plays an important
role in the secrecy performance, since it simultaneously affects the
rates of the legitimate channel and the eavesdropping channel.

\emph{Remark 3}: The derived ergodic secrecy rate can be extended to
the multi-cell case directly. In the scenario of multi-cell, both
the UEs and the Eves would suffer from the inter-cell interference
\cite{Multicell1,Multicell2}. Consequently, there exists the term of
inter-cell interference in the denominators of (\ref{eqn35}) and
(\ref{eqn36}). Similar to the inter-cluster interference, the power
of the inter-cell interference is regardless of $N_t$, since the
transmit beam is independent of the inter-cell interfering channels.

\emph{Remark 4}: As aforementioned, SIC at the UEs might be performed imperfectly, resulting in residual interference from the weaker UEs. In general, the residual interference can be modeled as a linear function of the power of the interfering signal, and the coefficient of imperfect SIC can be obtained by long-term measuring \cite{ImperfectSIC2, ImperfectSIC3}. Thus, the ergodic secrecy rate in the presence of imperfect SIC can be directly derived by adding the term of residual interference in the denominators of (\ref{eqn35}).

\emph{Remark 5}: In this paper, we consider that the Eve is or
pretends to be a UE. Thus, the BS is able to obtain full information
of the Eve, and derive the ergodic rate of the eavesdropping channel
in (\ref{eqn36}). However, if the BS has inexact information of
channel parameters, e.g., $\beta_m$ and $\rho_{m,0}$, there exist
uncertainty about the ergodic rate of the eavesdropping channel.
Consequently, the BS only has a possible range of the ergodic
secrecy rate, but not an exact value.

\subsection{Asymptotic Characteristics}
To provide clear insights into the impacts of the three key parameters of secure massive NOMA systems, namely $N_t$, $\rho_{m,n}$, and $P_{m,n}$, we carry out asymptotic analysis on the ergodic secrecy rate in two important scenarios.

\subsubsection{A Massive Number of BS Antennas}
We first reveal the impact of the number of BS antennas $N_t$ on the
ergodic secrecy rate $\bar{R}_{m,n}^{sec}$. As mentioned earlier,
the BS in 5G wireless networks is usually equipped with a
large-scale antenna array, and thus the number of BS antennas might
be very large. As $N_t$ approaches infinity, the inter-cluster
interference and the received noise are negligible compared to the
intra-cluster interference. Thereby, the ergodic rates of the
legitimate channel and the eavesdropping channel can be approximated as
\begin{eqnarray}
\bar{R}_{m,n}^{o}&\approx&\log_2\left(1+\frac{\alpha_{m,n}P_{m,n}\rho_{m,n}N_t}{\alpha_{m,n}\rho_{m,n}N_t\sum\limits_{i=1}^{n-1}P_{m,i}}\right)\nonumber\\
&\overset{N_t\rightarrow\infty}{=}&\log_2\left(1+\frac{P_{m,n}}{\sum\limits_{i=1}^{n-1}P_{m,i}}\right),\label{eqn37}
\end{eqnarray}
and
\begin{eqnarray}
\bar{R}_{m,n}^{e}&\approx&\log_2\left(1+\frac{\beta_{m}P_{m,n}\rho_{m,0}N_t}{\beta_{m}\rho_{m,0}N_t\sum\limits_{i=1}^{N_m}P_{m,i}}\right)\nonumber\\
&\overset{N_t\rightarrow\infty}{=}&\log_2\left(1+\frac{P_{m,n}}{\sum\limits_{i=1,i\neq n}^{N_m}P_{m,i}}\right).\label{eqn38}
\end{eqnarray}
In this scenario, the asymptotic ergodic secrecy rate can be expressed as
\begin{equation}
\bar{R}_{m,n}^{sec}=\log_2\left(1+\frac{P_{m,n}}{\sum\limits_{i=1}^{n-1}P_{m,i}}\right)-\log_2\left(1+\frac{P_{m,n}}{\sum\limits_{i=1,i\neq n}^{N_m}P_{m,i}}\right).\label{eqn39}
\end{equation}
It is found that there is a fixed gap between the rates of the
legitimate channel and the eavesdropping channel, which is caused by
the intra-cluster interference from the UEs with weaker channel
gains. In other words, the intra-cluster interference can be
exploited to confuse the eavesdropper, and thus improve the secrecy
performance. As an illustrative example, one can increase the power of the
weak user so as to increase the interference to the Eve, but does
not affect the received signal quality at the legitimate user.

\emph{Remark 6}: A massive number of BS antennas hardens the communication channels and both the inter-cluster interference and noise become negligible, resulting in the following phenomenons:
\begin{itemize}

\item The asymptotic ergodic secrecy rate is independent of the number of BS antennas, thus the ergodic secrecy rate will be saturated when the number of BS antennas is sufficiently large.

\item The asymptotic ergodic secrecy rate is independent of path loss and CSI accuracy, since the signals in the same cluster experience the same channel gain.

\item The asymptotic ergodic secrecy rate is independent of the BS transmit powers for the other clusters' signals, since the inter-cluster interference becomes asymptotically negligible.

\end{itemize}

\subsubsection{A High Transmit Power}
Then, we investigate the impact of BS transmit power on the ergodic secrecy rate. For convenience of analysis, we let $P_{m,n}=\nu_{m,n}P_{\mathrm{tot}}$, where $P_{\mathrm{tot}}$ is the total transmit power at the BS and $\nu_{m,n}$ is a nonnegative power allocation factor related to the UE$_{m,n}$ with a constraint $\sum\limits_{m=1}^{M}\sum\limits_{n=1}^{N_m}\nu_{m,n}=1$. As $P_{\mathrm{tot}}$ approaches infinity, the noise power at the receiver is negligible. Then, the ergodic rates of the legitimate channel and the eavesdropping channel can be simplified as
\begin{eqnarray}
\bar{R}_{m,n}^{o}&\approx&\log_2\left(1+\frac{\alpha_{m,n}P_{\mathrm{tot}}\nu_{m,n}\rho_{m,n}N_t}{\alpha_{m,n}\rho_{m,n}N_t\sum\limits_{i=1}^{n-1}P_{\mathrm{tot}}\nu_{m,i}+\alpha_{m,n}\sum\limits_{j=1,j\neq
m}^{M}\sum\limits_{i=1}^{N_j}P_{\mathrm{tot}}\nu_{j,i}}\right)\nonumber\\
&\overset{P_{\mathrm{tot}}\rightarrow\infty}{=}&\log_2\left(1+\frac{\nu_{m,n}\rho_{m,n}N_t}{\rho_{m,n}N_t\sum\limits_{i=1}^{n-1}\nu_{m,i}+\sum\limits_{j=1,j\neq
m}^{M}\sum\limits_{i=1}^{N_j}\nu_{j,i}}\right)\label{eqn40}
\end{eqnarray}
and
\begin{eqnarray}
\bar{R}_{m,n}^{e}&\approx&\log_2\left(1+\frac{\beta_{m}P_{\mathrm{tot}}\nu_{m,n}\rho_{m,0}N_t}{\beta_{m}\rho_{m,0}N_t\sum\limits_{i=1,i\neq n}^{N_m}P_{\mathrm{tot}}\nu_{m,i}+\beta_{m}\sum\limits_{j=1,j\neq
m}^{M}\sum\limits_{i=1}^{N_j}P_{\mathrm{tot}}\nu_{j,i}}\right)\nonumber\\
&\overset{P_{\mathrm{tot}}\rightarrow\infty}{=}&\log_2\left(1+\frac{\nu_{m,n}\rho_{m,0}N_t}{\rho_{m,0}N_t\sum\limits_{i=1,i\neq n}^{N_m}\nu_{m,i}+\sum\limits_{j=1,j\neq
m}^{M}\sum\limits_{i=1}^{N_j}\nu_{j,i}}\right),\label{eqn41}
\end{eqnarray}
respectively. Thus, the asymptotic ergodic secrecy rate in the scenario of a sufficiently high BS transmit power is given by
\begin{eqnarray}
\bar{R}_{m,n}^{sec}&=&\log_2\left(1+\frac{\nu_{m,n}\rho_{m,n}N_t}{\rho_{m,n}N_t\sum\limits_{i=1}^{n-1}\nu_{m,i}+\sum\limits_{j=1,j\neq
m}^{M}\sum\limits_{i=1}^{N_j}\nu_{j,i}}\right)\nonumber\\
&&-\log_2\left(1+\frac{\nu_{m,n}\rho_{m,0}N_t}{\rho_{m,0}N_t\sum\limits_{i=1,i\neq n}^{N_m}\nu_{m,i}+\sum\limits_{j=1,j\neq
m}^{M}\sum\limits_{i=1}^{N_j}\nu_{j,i}}\right).\label{eqn42}
\end{eqnarray}
It is seen from (\ref{eqn42}) that the UE$_{m,n}$ and the Eve$_m$ suffer the same inter-cluster interference, but the desired signal and the intra-cluster interference are different. In particular, the desired signal and the intra-cluster interference are functions of CSI accuracy and power allocation factors. Therefore, one can improve the ergodic secrecy rate by optimizing these two parameters, namely exploiting the inter-user interference during the stages of non-orthogonal channel estimation and non-orthogonal multiple access.

\emph{Remark 7}: In the high BS transmit power regime, we have the following observations:
\begin{itemize}

\item The asymptotic ergodic secrecy rate is independent of the BS transmit power, thus the ergodic secrecy rate will be saturated when the BS transmit power is sufficiently large.

\item The asymptotic ergodic secrecy rate is independent of path loss, since both the signal and the inter-user interference experience the same channel gain.

\item The asymptotic ergodic secrecy rate is independent of the BS power allocation for the other clusters' signals, since the inter-cluster interference is directly determined by the sum of the signal powers of the other clusters.

\end{itemize}

Furthermore, if the number of BS antennas is also very large, the inter-cluster interference is negligible. As a result, the asymptotic ergodic secrecy rate is only determined by the power allocation in a cluster. Hence, it makes sense to optimize the power allocation within a cluster for security enhancement. In what follows, we study the optimization of secrecy performance by exploiting the inter-user interference.

\section{Secrecy Performance Optimization of Massive NOMA}
An advantage of the proposed secure massive NOMA scheme is that the inter-user interference can be leveraged to confuse the Eves, and thus enhance the secrecy performance. To fully exploit the benefits of inter-user interference for security provisioning, it is desired to optimize the system parameters according to channel conditions. In particular, the inter-user interference is determined by non-orthogonal channel estimation and non-orthogonal multiple access, it is likely to optimize these two important stages of secure massive NOMA systems. Inspired by that, we propose to optimize these two stages from the perspectives of the maximization of the secrecy rate and the minimization of the transmit power, respectively.

\subsection{Exploiting Inter-User Interference in Non-Orthogonal Channel Estimation}
As analyzed above, the Eves send the same training sequences as the legitimate users to interfere with the channel estimation, resulting in the information leakage during the stage of information transmission. According to (\ref{eqn36}), the amount of information leakage is directly determined by the CSI accuracy of the eavesdropper channel. Since the CSI accuracy of the eavesdropper channel is also affected by the training sequences of the UEs, it is possible to coordinate the transmit powers of the UEs to combat the interception of the Eves.

\subsubsection{The Maximization of Secrecy Rate}
First, we optimize the non-orthogonal channel estimation from the perspective of maximizing the secrecy rate. However, the secrecy rate, defined as the difference of the legitimate channel rate and the eavesdropping channel rate, is in general non-convex. To improve tractability of this problem, we transform it as the maximization of the minimum legitimate channel rate subject to a constraint on the maximum eavesdropping channel rate. This problem formulation ensures a rate gap to facilitate secure communication. Furthermore, one can adjust the rate gap for meeting the requirements of various secure communications. In this case, the optimization of non-orthogonal channel estimation can be described as the following optimization problem:
\begin{eqnarray}
\mathrm{OP1}\!\!\!\!&:&\!\!\!\!\max\limits_{\mathbf{Q}}\min\limits_{m,n}\bar{R}_{m,n}^o\nonumber\\
\textrm{s.t. C1}\!\!\!\!&:&\!\!\!\!\bar{R}_{m,n}^e\leq r^e,\nonumber\\
\textrm{C2}\!\!\!\!&:&\!\!\!\!Q_{m,n}\leq Q_{m,n}^{\max},\forall m,n,
\end{eqnarray}
where $\mathbf{Q}$ is a collection of the transmit powers of the UEs for channel estimation, and $Q_{m,n}^{\max}$ is the maximum transmit power of the UE$_{m,n}$. Constant $r^e$ is the maximum tolerable ergodic rate achieved at the Eves, which is also used to ensure the rate of the eavesdropping channel less than the transmit rate such that the Eves cannot perform SIC perfectly. To solve such a maxmin problem, OP1 is first transformed as the following equivalent optimization problem
\begin{eqnarray}
\mathrm{OP2}\!\!\!\!&:&\!\!\!\!\max\limits_{\mathbf{Q}, r^o}r^o\nonumber\\
\textrm{s.t. C1,}\!\!\!\!\!&&\!\!\!\!\!\mathrm{C2},\nonumber\\
\textrm{C3}\!\!\!\!&:&\!\!\!\!\bar{R}_{m,n}^o\geq r^o,\forall m,n,
\end{eqnarray}
where $r^o$ is a minimum ergodic rate of the UEs. According to the expression of $\bar{R}_{e}$ in (\ref{eqn36}), C1 can be rewritten as
\begin{eqnarray}
\beta_m^2P_{m,n}N_t\tau Q_{m,0}\!\!\!\!&\leq&\!\!\!\!\left(2^{r_e}-1\right)\beta_m^2\sum\limits_{i=1,i\neq n}^{N_m}P_{m,i}N_t\tau Q_{m,0}\nonumber\\
&&\!\!\!\!+\left(2^{r_e}-1\right)\left(\beta_m\sum\limits_{j=1,j\neq m}^{N_m}\sum\limits_{i=1}^{N_j}P_{j,i}+1\right)\left(1+\sum\limits_{k=0}^{N_m}\alpha_{m,k}Q_{m,k}\tau\right).\label{eqn43}
\end{eqnarray}
Note that (\ref{eqn43}) is a linear inequality of $Q_{m,n}$. Similarly, C3 also can be transformed as a linear inequality related to $Q_{m,n}$. Thus, for a given $r^o$, OP2 becomes a standard linear programming problem, which can be easily solved by some existing optimization softwares. Then, through a one-dimensional search on $r^o$, it is possible to obtain the optimal power control scheme about $\mathbf{Q}$. The algorithm for maximizing the secrecy rate by exploiting the inter-user interference in non-orthogonal channel estimation is summarized as Algorithm 1 at the top of the next page.
\begin{algorithm}
\caption{: Power control for maximizing the secrecy rate}
\label{alg1}
\begin{algorithmic}
\STATE{\textbf{Step 1:}\quad Initialize the parameters by letting $r^o=r^e$;}
\STATE{\textbf{Step 2:}\quad Solve the optimization problem OP2, and obtain $Q_{m,n}, \forall m,n$;}
\STATE{\textbf{Step 3:}\quad Update $r^o$ by letting $r^o=r^o+\Delta_o$, where $\Delta_o$ is a small positive real number. If OP2 is solvable, then go to Step 2.}
\STATE{\textbf{Step 4:}\quad Output $Q_{m,n}, \forall m,n$ with the largest $r^o$.}
\end{algorithmic}
\end{algorithm}

\subsubsection{The Minimization of Total Power Consumption}
Since the energy at the UE is limited, it is desired to find the minimum power to achieve a given objective of ergodic secrecy rate. In the case, the optimization problem can be formulated as
\begin{eqnarray}
\mathrm{OP3}\!\!\!\!&:&\!\!\!\!\min\limits_{\mathbf{Q}}\sum\limits_{j=1}^{M}\sum\limits_{i=1}^{N_j}Q_{j,i}\nonumber\\
\textrm{s.t. C1,}\!\!\!\!\!&&\!\!\!\!\!\textrm{C2, C3}.
\end{eqnarray}
As mentioned above, constraints C1 and C3 can be transformed as linear inequalities related to $Q_{m,n}$. Thus, OP3 is also a linear programming problem, and can be solved directly by using some optimization softwares, e.g. CVX \cite{CVX}.

\subsection{Exploiting Inter-User Interference in Non-Orthogonal Multiple Access}
It is noted that the inter-user interference in non-orthogonal channel estimation can only be used to confuse the Eve in the same cluster, but the inter-user interference in non-orthogonal multiple access affects all the Eves. Therefore, the optimization of non-orthogonal multiple access is more beneficial to the enhancement of the secrecy performance. Similarly, we also exploit the benefits of inter-user interference from the perspectives of maximizing the secrecy rate and minimizing the total power consumption, respectively.

\subsubsection{The Maximization of Secrecy Rate}
To design a computation-efficient solution, we transform the maximization of secrecy rate as the maximization of the legitimate channel rate subject to a constraint on the eavesdropping channel rate. Thus, the optimization problem for maximizing the secrecy rate by exploiting the inter-user interference in non-orthogonal multiple access can be formulated as
\begin{eqnarray}
\mathrm{OP4}\!\!\!\!&:&\!\!\!\!\max\limits_{\mathbf{P}, r^o}r^o\nonumber\\
\textrm{s.t. C1,}\!\!\!\!\!&&\!\!\!\!\!\mathrm{C3}\nonumber\\
\textrm{C4}\!\!\!\!&:&\!\!\!\!\sum\limits_{j=1}^{M}\sum\limits_{i=1}^{N_j}P_{j,i}\leq P_{\textrm{tot}}\nonumber\\
\textrm{C5}\!\!\!\!&:&\!\!\!\!P_{m,n}\leq P_{m,n+1},\forall m,n,
\end{eqnarray}
where $\mathbf{P}$ is the collection of the transmit powers of the BS for information transmission and C5 is imposed to facilitate SIC at the UEs. It is easy to rewrite C1 and C3 as linear inequalities of the transmit powers of the BS $P_{m,n}$. Therefore, one can obtain the solution to OP4 for a given $r^o$. Through a one-dimension search on $r^o$, it is possible to derive the optimal power allocation scheme at the BS. In specific, the algorithm for maximizing the secrecy rate by exploiting the inter-user interference in non-orthogonal multiple access can be summarized as
\begin{algorithm}
\caption{: Power allocation for maximizing the secrecy rate}
\label{alg1}
\begin{algorithmic}
\STATE{\textbf{Step 1:}\quad Initialize the parameters by letting $r^o=r^e$;}
\STATE{\textbf{Step 2:}\quad Solve the optimization problem OP4, and obtain $P_{m,n}, \forall m,n$;}
\STATE{\textbf{Step 3:}\quad Update $r^o$ by letting $r^o=r^o+\Delta_o$, where $\Delta_o$ is a small positive real number. If OP4 is solvable, then go to Step 2.}
\STATE{\textbf{Step 4:}\quad Output $P_{m,n}, \forall m,n$ with the largest $r^o$.}
\end{algorithmic}
\end{algorithm}

\subsubsection{The Minimization of Total Power Consumption}
Then, we also consider the optimization of non-orthogonal multiple access from the perspective of minimizing the power consumption at the BS. In this case, the power allocation can be described as the following optimization problem:
\begin{eqnarray}
\mathrm{OP5}\!\!\!\!&:&\!\!\!\!\min\limits_{\mathbf{P}}\sum\limits_{j=1}^{M}\sum\limits_{i=1}^{N_j}P_{j,i}\nonumber\\
\textrm{s.t. C1,}\!\!\!\!\!&&\!\!\!\!\!\textrm{C3, C4, C5}.
\end{eqnarray}
Note that C1 and C3 can be transformed as linear inequalities related to $P_{j,i}$, and C4 and C5 are linear constraints. Thus, OP5 is a linear programming problem, which can be easily solved by some optimization softwares.

The proposed algorithms are also applicable in the case of
multiple-cell and imperfect SIC. This is because the inter-cell
interference in the multiple-cell scenario is equivalent to the
inter-cluster interference, and the residual interference caused by
imperfect SIC is similar to the intra-cluster interference.
Moreover, in the presence of uncertainty about ergodic secrecy rate,
the proposed algorithms can be redesigned as robust ones by through
a one-dimensional searching over the uncertain region of the ergodic
secrecy rate.

\section{Numerical Results}

To evaluate the performance of the proposed secure massive NOMA
scheme, we present several simulation results in different
scenarios. Without extra specification, we set $N_t=64$, $M=12$, and
$N_m=4, \forall m$. It is assumed that all the UEs have the same
maximum transmit power for channel estimation. For convenience, we
use PSNR to denote the ratio of the total BS transmit power and the
noise power (in dB), QSNR to denote the ratio of the maximum UE
transmit power and the noise power (in dB), and USNR to denote the
ratio of the Eve transmit power and the noise power (in dB),
respectively.

\begin{figure}[h]
\centering
\includegraphics [width=0.6\textwidth] {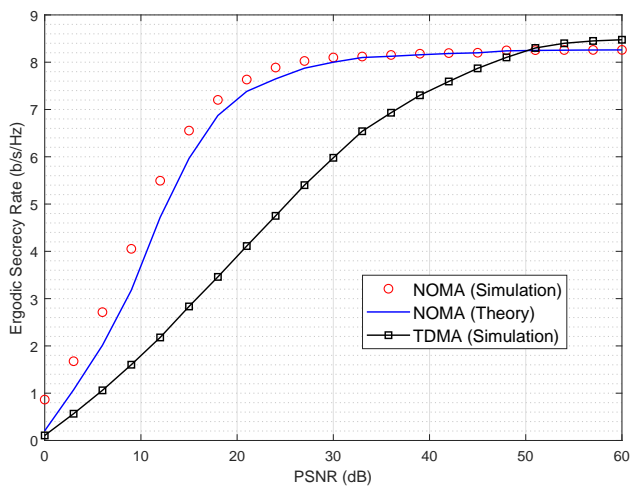}
\caption {The accuracy of theoretical analysis and performance comparison of NOMA and TMDA schemes.} \label{Fig2}
\end{figure}

Firstly, we check the accuracy of the derived theoretical expression
for ergodic secrecy rate. As seen in Fig. \ref{Fig2}, the
theoretical expression is a tight lower bound of the simulation
results, especially in the high PSNR regime. Both theoretical
analysis and simulation results show that with the increase of the
PSNR, the ergodic secrecy rate first improves and then saturates
when the PSNR is sufficiently large. This is consistent with the
asymptotic characteristics of a high BS transmit power analyzed in
Section III-B. Moreover, we compare the NOMA scheme and a TDMA
scheme in the sense of sum ergodic secrecy rate. Note that we assume
that the TDMA scheme has full CSI at the BS. However, it is found
that the NOMA scheme achieves a superior performance compared to the
TDMA scheme in low and medium PSNR regions. This is because each UE
based on the TDMA scheme only occupies a small fraction of time
duration, resulting in a low spectral efficiency. At very high PSNR,
the TDMA scheme performs slightly better than the NOMA scheme. The
reason that the interference is dominant at very high PSNR, but the
TDMA scheme is free of intra-cluster interference and inter-cluster
interference. However, the BS in general operates in low and medium
PSNR regions. Therefore, the proposed NOMA scheme can effectively
guarantee secure communication of 5G wireless networks.

%
%Firstly, we validate the accuracy of the derived theoretical lower bound on the spectral efficiency. As seen in Fig. \ref{Fig2}, in the whole BSNR region, the theoretical lower bound has the same trend as the simulation results, and the gap between theoretical and simulation results is slight. Especially when the number of BS antennas is sufficiently large, the theoretical bound well coincides with the simulation results. Moreover, it is found that the spectral efficiency has two characteristics. On on hand, the spectral efficiency improves as the number of BS antennas increases, but the gain by adding the same number of BS antennas decreases gradually. On the other hand, the spectral efficiency improves as BSNR increases, but it will be saturated when BSNR is high enough. In particular, the saturation values in the scenarios of different numbers of BS antennas are equal.

\begin{figure}[h]
\centering
\includegraphics [width=0.6\textwidth] {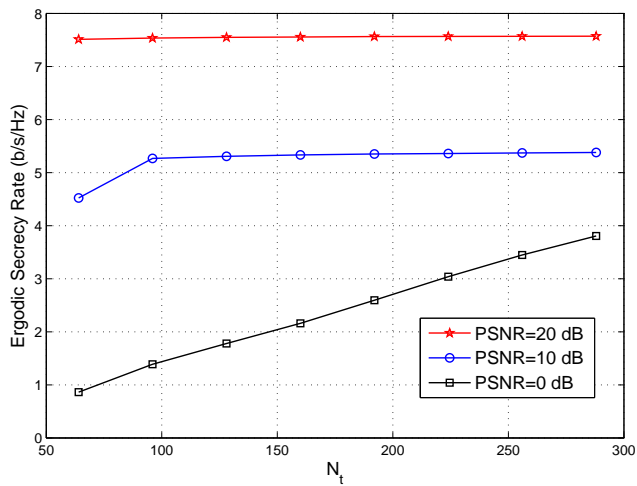}
\caption {The effect of the number of BS antennas on the ergodic secrecy rate.} \label{Fig3}
\end{figure}

Secondly, we show the effect of the number of BS antennas on the sum
of ergodic secrecy rate, cf. Fig. \ref{Fig3}. It is found that at
low PSNR, as the number of BS antennas increases, the sum of ergodic
secrecy rate improves sharply. Thus, one can support secure massive
access by simply increasing the number of BS antennas, which is a
major advantage of massive MIMO systems with NOMA. As theoretically
analyzed, the ergodic secrecy rate asymptotically approaches a
constant as the number of BS antennas increases. Especially, if
both the number of BS antennas and the BS transmit powers are sufficiently large,
the performance gain achieved by increasing the BS transmit power is quite
limited. This is because the ergodic secrecy rate is independent of
the BS transmit power under such a condition which reconfirms the
theoretical analysis.

%
%Secondly, we show the impact of the number of BS antennas on the spectral efficiency. It is seen in Fig. \ref{Fig3} that the spectral efficiencies with different BSNRs sharply improves by adding the number of BS antennas, which confirms the advantages of the large-scale antenna array in supporting massive access. However, the spectral efficiency will be saturated when the number of BS antennas is sufficiently large. Thus, given channel conditions, it is not necessary to utilize a very large number of BS antennas from the perspective of maximizing the spectral efficiency.
%
\begin{figure}[h]
\centering
\includegraphics [width=0.6\textwidth] {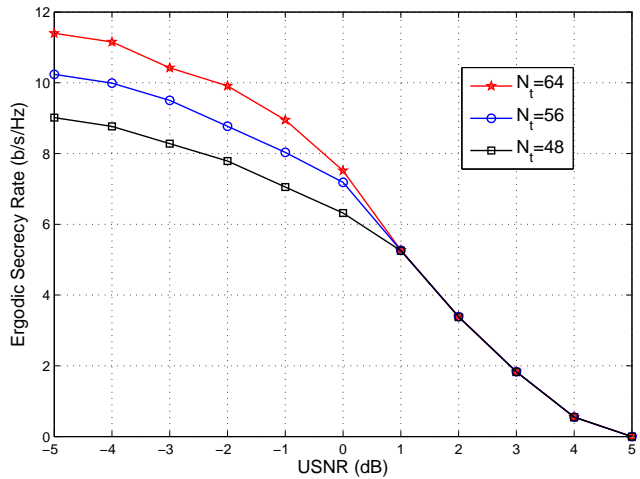}
\caption {The impact of the Eve transmit power on the ergodic secrecy rate.} \label{Fig4}
\end{figure}

Thirdly, in Fig. \ref{Fig4}, we investigate the impact of the Eve transmit power on the
sum of the ergodic secrecy rate for a given QSNR=-5 dB. It is
intuitive that the sum of ergodic secrecy rate decreases as the Eve
transmit power increases, since the CSI accuracy of the legitimate
channel decreases. However, by exploiting the inter-user
interference in non-orthogonal channel estimation, the effect of the
active attacking from the Eve is limited. As seen in Fig.
\ref{Fig4}, even with a relative high attacking power, e.g., USNR=0
dB, the proposed scheme can achieve a high ergodic secrecy rate.
Especially, if USNR is not so high compared to QSNR, it is possible to
enhance the secrecy performance by adding more BS antennas,
and thus alleviate the impact of the active attacking from the Eve on the secrecy performance.

%
%Fig. \ref{Fig4} investigates the effect of relay transmit powers on the spectral efficiency. It is worthy pointing out that the relays utilize the same maximum power to forward the signals. As analyzed in Section III.B, the spectral efficiency will be saturated as the RSNR increases. In this scenario, it is possible to further improve the spectral efficiency by adding the number of BS antennas. However, the gain by adding the same number of BS antennas decreases gradually. This is because the spectral efficiency will also be saturated as the number of BS antennas increases as shown in Fig. \ref{Fig3}.
%
\begin{figure}[h]
\centering
\includegraphics [width=0.6\textwidth] {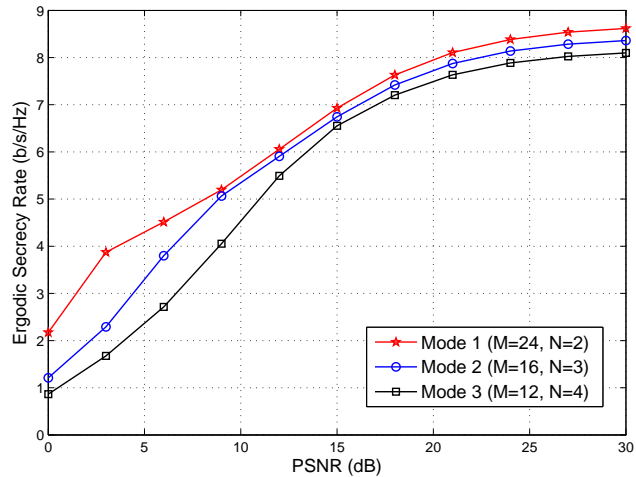}
\caption {Performance comparison with different modes of user clustering.} \label{Fig5}
\end{figure}

Then, we compare the sum of ergodic secrecy rates with different
modes of user clustering. In particular, for a given total number of
the UEs, $K=48$, the UEs can be partitioned into different numbers
of clusters. In this case, we consider three modes with 24, 16, and
12 clusters, respectively. Note that we utilize the pilot sequences
of fixed length for all modes. As shown in Fig. \ref{Fig5}, the sum
of ergodic secrecy rate improves as the number of user clusters
increases (equivalently the number of UEs in a cluster decreases).
This is because the CSI accuracy improves if the number of UEs in a
cluster decreases. However, increasing the number of clusters leads
to longer training sequences. Moreover, a smaller number of UEs in a
cluster is not capable of supporting massive access. Thus, it is
desired to select a proper mode to balance the sum of ergodic
secrecy rate and the length of the training sequence.
%
%Then, we compare the spectral efficiencies of power allocation schemes at the BS. In Fig. \ref{Fig5}, we take equal power allocation and fixed power allocation as two baseline schemes. Note that equal power allocation distributes the BS transmit powers to all UEs equally, and fixed power allocation proposed in \cite{NOMA2} distributes the power $\frac{n}{\sum\limits_{k=1}^{K_m}k}\frac{P_{\mathrm{tot}}}{M}$ to the $n$th UEs related to the $m$th relay. It is found that equal power allocation performs better than fixed power allocation, but the latter facilitates SIC at the UE. This is because fixed power allocation distributes more powers to the UE with a weaker channel gain. The proposed power allocation scheme can achieve the best performance, while it also facilitates SIC due to the constraint condition C3. Thus, the proposed power allocation scheme can significantly improve the performance of massive access.
%
\begin{figure}[h]
\centering
\includegraphics [width=0.6\textwidth] {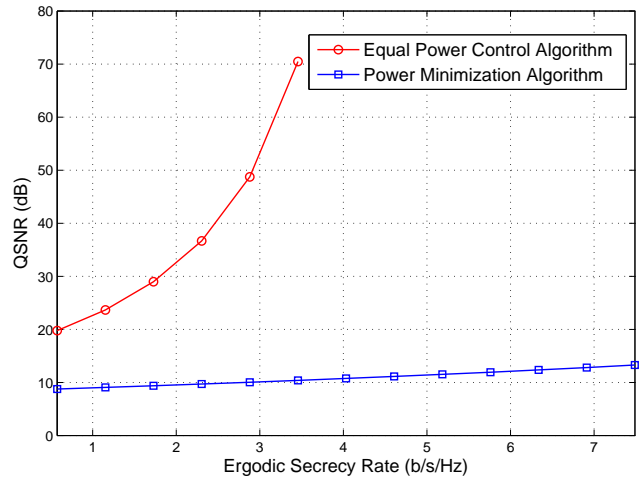}
\caption {Performance comparison with different power control algorithms.} \label{Fig6}
\end{figure}

Next, we show the advantage of the proposed power control algorithm for exploiting the benefits of inter-user interference at the stage of non-orthogonal channel estimation over an equal power control algorithm. To be specific, under the equal power control algorithm, all the UEs use the same power to send the training sequences. As seen in Fig. \ref{Fig6}, given the requirement of ergodic secrecy rate, the proposed power minimization algorithm has a very low total power consumption. Especially, as the required ergodic secrecy rate increases, the proposed algorithm needs only a slight increment of the power, but the power consumption of the equal power control algorithm increases sharply. This is because the UEs have different path loss and the equal power control leads to a low power utilization efficiency. Thus, the proposed algorithm can effectively exploit the benefits of inter-user interference with a small power consumption.
%
%Next, we show the advantage of the proposed power control scheme at the relays over a fixed power control scheme. To be specific, the fixed power control scheme uses up the power budge at the relays. As seen in Fig. \ref{Fig5}, fixed power control leads to a large performance loss. This is because a high relay transmit power may amplify the forwarding noise and increase the inter-relay interference. The proposed power control scheme is capable of achieving a balance between enhancing the desired signal and mitigating the interference and noise, and thus improves the spectral efficiency.
%
%
\begin{figure}[h]
\centering
\includegraphics [width=0.6\textwidth] {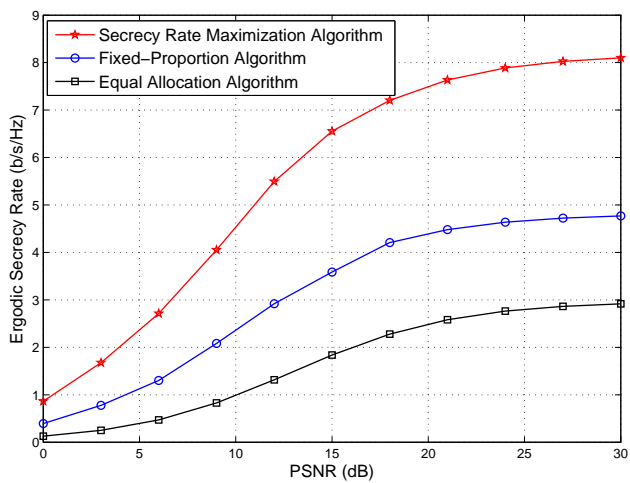}
\caption {Performance comparison with different power allocation algorithms.} \label{Fig7}
\end{figure}

Finally, we compare the proposed ergodic secrecy rate maximization power allocation algorithm with two baseline algorithms, namely a fixed-proportion power allocation algorithm and an equal power allocation algorithm. As the name implies, the fixed-proportion power allocation algorithm proposed in \cite{NOMA2} allocates $\frac{n}{\sum\limits_{k=1}^{N_m}k}\frac{P_{\mathrm{tot}}}{M}$ to the UE$_{m,n}$, and the equal power allocation algorithm distributes the BS transmit power to all UEs equally. It is found in Fig. \ref{Fig7} that the proposed algorithm performs the best in the whole PSNR region. Especially, as PSNR increases, the performance gain with respect to the two baseline algorithms becomes larger. This is because the proposed algorithm adaptively adjusts the power allocation according to channel conditions and system parameters. Hence, at the stage of multiple access, the proposed algorithm also can effective exploit the benefits of inter-user interference to enhance the security of 5G wireless networks with massive connections.

%
%Finally, we validate the effectiveness of the proposed multiple-relay aided NOMA technique in supporting massive access compared to the traditional space division multiple access (SDMA) and time division multiple access (TDMA) techniques. Specifically speaking, SDMA allows $M$ UEs to access the spectrum via $M$ different relays in a time slot, while TDMA only makes one UE to communicate with the BS via a relay in a time slot. Although SDMA and TDMA can decrease and even avoid the interference, there exits a large performance gap with respect to the NOMA. Especially for TDMA, as shown in Fig. \ref{Fig7}, there is a great performance loss. This is because TDMA utilizes much more time slots to complete the access of a massive number of UEs. Thus, the proposed multiple-relay aided NOMA technique is appealing and effective in supporting massive access.

\section{Conclusion}
This paper addressed the security issue in 5G wireless networks with massive access by exploiting the inter-user interference. We first designed a fully non-orthogonal communication framework including channel estimation and multiple access according to the characteristics of massive access. Then, we revealed the impact of inter-user interference on the secrecy performance. Finally, we proposed two optimization schemes for channel estimation and multiple access to fully exploit the benefits of inter-user interference for enhancing the security of 5G wireless networks.

\end{spacing}
\end{document}